# Towards a dynamics-based estimate of the extent of HR 8799's unresolved warm debris belt

Bruna Contro[1], Rob Wittenmyer[1,2,3], Jonti Horner[2,3], Jonathan P. Marshall[1,3]

[1]School of Physics, UNSW Australia, Sydney, New South Wales 2052, Australia
[2]Computational Engineering and Science Research Centre, University of Southern Queensland, Toowoomba, Queensland 4350, Australia
[3]Australian Centre for Astrobiology, UNSW Australia, Sydney, New South Wales 2052, Australia

**Summary:** In many ways, the HR 8799 planetary system resembles our Solar system more closely than any other discovered to date – albeit significantly younger, and on a larger and more dramatic scale. The system features four giant planets and two debris belts. The first of these belts lies beyond the orbit of the outermost planet, and mirrors the location of our Solar system's Edgeworth-Kuiper belt. The second, which has yet to be fully observationally characterised, lies interior to the orbit of the innermost known planet, HR8799 e, and is an analogue to our Asteroid Belt. With such a similar architecture, the system is a valuable laboratory for examining exoplanetary dynamics, and the interaction between debris disks and giant planets.

In recent years, significant progress has been made in the characterisation of the outer of HR8799's debris disks, primarily using the Herschel Space Observatory. In contrast, the inner disk, which lies too close to its host star to be spatially resolved by that instrument, remains poorly understood. This, in turn, leaves significant questions over both the location of the planetesimals responsible for producing the observed dust, and the physical properties of those grains.

We have performed the most extensive simulations to date of the inner, unresolved debris belt around HR 8799, using UNSW Australia's supercomputing facility, *Katana*. In this work, we present the results of integrations following the evolution of a belt of dynamically hot debris interior to the innermost planet, HR8799 e, for a period of 60 million years, using an initial population of 500,000 massless test particles. These simulations have enabled the characterisation of the extent and structure of the inner belt, revealing that its outer edge must lie interior to the 3:1 mean-motion resonance with HR8799 e, at approximately 7.5 au, and highlighting the presence of fine structure analogous to the Solar system's Kirkwood gaps. In the future, our results will also allow us to calculate a first estimate of the small-body impact rate and water delivery prospects for any potential terrestrial planet(s) that might lurk, undetected, in the inner system.

# Introduction

In the past few years, the search for planets around other stars has moved from focussing on the biggest and easiest planets to find (e.g. [1][2][3]), to looking for planets more like the Earth. As part of this process, the search for Solar system analogues has come to the fore, for a number of reasons. The first is purely practical – radial velocity observations of sun-like stars can only detect planets with orbital periods shorter than the time a given star has been targeted with observations. In order to discover a Jupiter-like planet, then, a given star must have been observed on a decadal timescale. For this reason, it is only in the last few years that the first true Jupiter-analogues have been found (e.g. [4][5][6]). Furthermore, the discovery of smaller, more Earth-like planets, has required the development of more precise tools, as well as larger observational datasets. Once again, that search has only recently begun to bear fruit (e.g. [7][8][9]).

In this push towards finding planetary systems that resemble our own, different techniques have brought complementary results to the table. The search for the smallest, most Earth-like planets, has been led by the *Kepler* spacecraft, using the transit technique (e.g. [7][10]). At the same time, the search for Jupiter-analogues has been led by direct imaging and radial velocity programs, which continue to find massive planets at ever increasing distances from their host stars.

Beyond the simple practical reasons for the slow onset of the search for Solar system analogues, an additional and key motivation for that search is that finding planetary systems that truly resemble our own is critically important, particularly since it enables us to gain a better understanding of our own Solar system and its place in the universe.

Despite the rapid progress made over the past two decades, technological limitations have, to date, resulted in very few exoplanetary systems being discovered that truly resemble our own. Possibly the prime example of such a system is that orbiting the star HR8799. The discovery of the first three planets in that system was announced in 2008 [11], using direct images of the system taken with the Keck and Gemini observatories. A fourth giant planet was found in 2010 [12] as a result of follow-up observations. With its four giant planets [12] and a circumstellar disk [13] comprised of two debris belts [14], this system stands as the most "Solar system like" of all that have been discovered to date, and as such is an important laboratory that will allow us to greatly improve our understanding of the formation, evolution, and scarcity of planetary systems like our own.

HR 8799 is a young A-type star, located ~ 40 pc from the Earth [15]. Whilst its age remains, to date, somewhat poorly determined, its relative youth is well accepted. Whilst [16] suggest that HR 8799 is no older than 50 million years (Myr), [17] suggest an age of 30 Myr, based on the stars likely membership of the Columba Association. When announcing the discovery of HR8799 e, [12] take a conservative perspective on the star's age. They consider two different ages for the planetary system – 30 Myr, based on the star's membership of the Columba Association, and 60 Myr, following previous work [11]. See Table 1 for a summary of the stellar parameters used in this work.

*Table 1. Stellar Parameters for HR 8799.*

| Parameter | Value | Reference |
|---|---|---|
| Age (Myr) | 30 | [17] |
|  | 60 | [11] |
| Teff (K) | 7193 ± 87 | [18] |
| L (L$_\odot$) | 5.05 ± 0.29 | [18] |
| M (M$_\odot$) | 1.56 | [12][19] |

In Table 2, we present the best-fit orbital parameters of the planets, as presented in a recent dynamical study [19]. For this work we considered the primary values only, ignoring the errors. A schematic showing the orbits of the four planets in the HR 8799 system is shown in Figure 1. In this resonant architecture (planets trapped in a double Laplace resonance with orbital period ratios of 1:2:4:8, as described in [19]), the orbits of the planets remain dynamically stable on long timescales.

*Table 2. The orbital elements of the four planets orbiting the star HR 8799, as used in our integrations. These elements are those of the double-Laplace resonant architecture proposed in Table 1 of [19]. Here, m is the mass of the planet, in Jupiter masses, a is the orbital semi-major axis, in au, and e is the eccentricity of the orbit. i is the inclination of the orbit to the plane of the sky, and Ω, ω and M are the longitude of the ascending node, the longitude of pericentre, and the mean anomaly at the epoch 1998.83, respectively.*

|  | $m$ [m$_{jup}$] | $a$ [au] | e | $i$ [deg] | Ω [deg] | ω [deg] | $M$ [deg] |
|---|---|---|---|---|---|---|---|
| HR 8799 e | 9±2 | 15.4±0.2 | 0.13±0.03 | 25±3 | 64±3 | 176±3 | 326±5 |
| HR 8799 d | 9±2 | 25.4±0.3 | 0.12±0.02 |  |  | 91±3 | 58±3 |
| HR 8799 c | 9±2 | 39.4±0.3 | 0.05±0.02 |  |  | 151±6 | 148±6 |
| HR 8799 b | 7±2 | 69.1±0.2 | 0.020±0.003 |  |  | 95±10 | 321±10 |

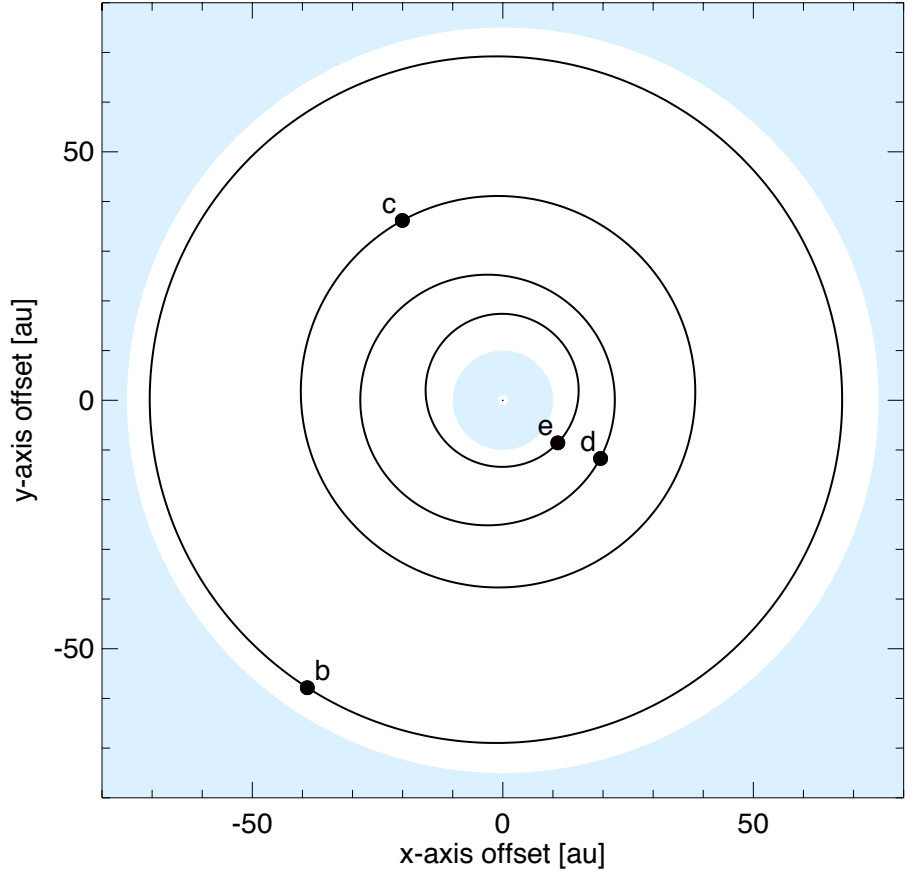

*Figure 1. This is a schematic plot of the HR 8799 system. The four planets are at their correct orbital separations from the star, and location around the star, but are presented on circular rather than eccentric orbits. The light blue regions are those parts of the system where circumstellar debris is expected to reside in order to explain the observed infrared excess described in Matthews et al., 2014.*

In addition to the four planets detected by [12], the HR 8799 system also hosts two debris belts and a halo of small grains and dust out to > 1000 au [14]. Whilst the outer belt has been spatially resolved by [13], [16] and [14], and shown to extend from 100 (±10 au) to 310 au, the inner, warm belt is, to date, poorly understood. The best estimate, based on a black body fit to the observed dust temperature ([20][21]), is that it must lie between 1 and 10 au. This information, coupled with our knowledge of the Keplerian orbital elements for HR 8799 b, c, d and e, gives us enough information to, for the first time, attempt to constrain its extent and interior structure using *n*-body simulations.

## Material and Methods

We performed the most extensive simulations to date of the inner debris belt of HR 8799 using the numerical integrator package MERCURY [22], using UNSW's *Katana* supercomputing facility, and building on our earlier work studying the system [23]. The disk, initially composed of 500,000 massless test particles, was followed for 60 Myr (in order to ensure compatibility with the ages proposed by [11] and [12]) to trace its evolution, with an integration time-step of 7 days (approximately 1/2500th of the orbital period of HR8799 e). Each test particle was followed (i) until it was ejected from the system (upon reaching a barycentric distance of 1000 au); (ii) impacted on one of the giant planets; or (iii) collided with the central body.

For this work, each of the six orbital elements for the test particles were randomly allocated to lie within a set range. The semi-major axes were chosen to lie between 1 and 10 au (with the inner and outer edges chosen in accordance with [24] and [14] descriptions of the inner disk). Once the semi-major axes were set, the eccentricity of the massless test particles was allocated to lie between 0.1 and 1.0[1]. Each of the test particles was then given an orbital inclination between 0 and 25 degrees (a range similar to that occupied by the great majority of objects in the Solar system's Asteroid and Edgeworth-Kuiper belts). The rotational orbital elements for each particle were each separately randomly dispersed between 0 and 360 degrees. The result of these processes was a dynamically excited disk of debris that filled the inner reaches of the HR 8799 planetary system, with orbital elements within the ranges detailed in Table 3. The orbital elements of the particles, as well as those of the four giant planets integrated, were output at 6 Myr intervals, allowing us to trace the evolution of the warm belt over the lifetime of the system.

*Table 3: The range of values covered by the population of test particles considered in this work, at the start of the integrations. Within these values, the orbital elements of each test particle were randomly generated.*

| Orbital element | Range of values considered |
|---|---|
| | |
| Semi-major axis (au) | $1 < a < 10$ |
| Eccentricity | $0.1 < e < 1.0$ |
| Inclination (°) | $0 < i < 25$ |
| Argument of Periastron (°) | $0 < \omega < 360$ |
| Longitude of the Ascending Node (°) | $0 < \Omega < 360$ |
| Mean Anomaly (°) | $0 < M < 360$ |
| | |
| *Subsidiary considerations* | |
| Periastron distance (au) | $q > 0.1$ |

For this work, we followed the best-fit parameters for HR 8799 b, c, d and e and stellar mass of $1.56 M_\odot$ described by [19], as shown in Table 2. In addition, the location of the classical "Habitable Zone" (HZ; the region around the star in which an Earth like planet might reasonably be expected to be able to host liquid water on its surface, e.g. [25]) was estimated to be 1.974 (inner edge) to 3.407 au (outer edge), following [26] and [27], using the stellar parameters shown in Table 1.

We note that, in this paper, we consider the inner debris disk as a distribution of massless test particles. These particles individually represent a planetesimal inserted into the HR8799 planetary system on the given orbit. This approach neglects radiative forces and the perturbing influence of a massive asteroid belt on the orbits of the planets.

---

[1] These simulations are constructed to complement those presented in [34], where we considered a dynamically cold, unexcited disk, and are intended primarily to determine the extremes at which the disk loses stability – hence the very broad range in orbital eccentricity considered. In practice, the highest eccentricity of any test particle in our initial sample was 0.992793 – a result of an additional constraint that no test particle move on an orbit with initial periastron less than 0.1 au.

If we had used massive particles, the total mass of the debris disk would have become a limiting factor in the simulations, pulling on the planets and altering their orbits (as seen in simulations of a massive outer debris disk for HR8799 [33]). As such, these simulations are limited in relevance to the case where the inner disk is much less massive than the innermost planet. This is expected to be the case, since the collisional timescale for several tens of Earth masses of material at ~ 5-10 au would be << 60 Myr, causing a rapid depletion and removal of the inner disk.

By neglecting radiative forces we can only comment on the regions within which planetesimals would reside. Dust grains, which migrate under the influence of radiative forces, spread beyond their formation region both inward and outward through the system, sometimes becoming trapped in resonances with planets. As such the distribution of dust grains would not directly mirror that of their parent planetesimals.

The results of these simulations, representing a simplification of the true system's dynamical interactions, still provide a first glimpse of the extent, structure and complexity of the inner system.

## Results

The instantaneous semi-major axes and eccentricities of the test particles in the inner debris disk of HR 8799 are shown as a function of time in Figure 2. Panel one shows the initial orbits of the 500,000 massless test particles, whilst the other panels show the elements of those particles surviving at each of the time-steps in question, in which each black dot represents a surviving test particle. Figure 3 shows in more detail the final distribution of the surviving test particles after the full 60 Myr have elapsed as well as the location of the HZ (green bar) and a few select mean-motion resonances with HR 8799 e (red lines), whilst Figure 4 shows the number of test particles that remain in the simulation as a function of time in our simulations.

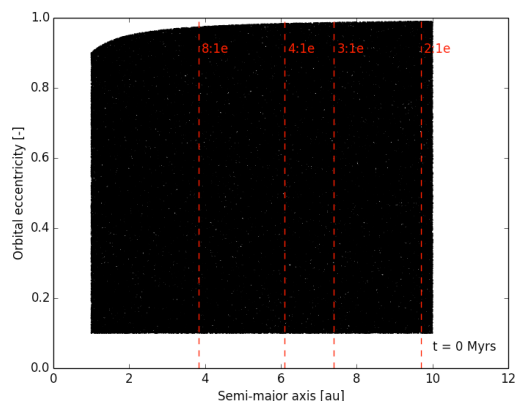

**Panel 1:** 0 Myr.

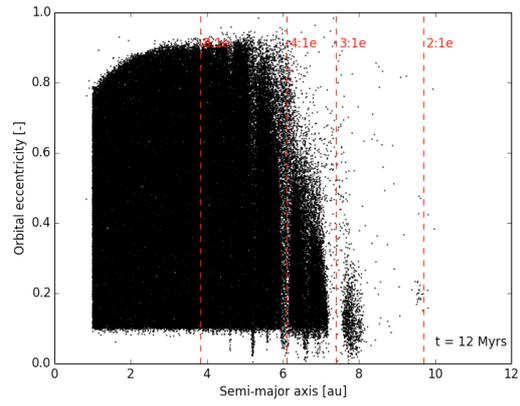
**Panel 2:** 12 Myr.

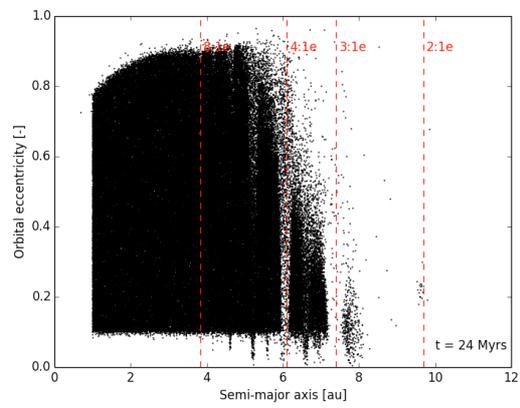
**Panel 3:** 24 Myr.

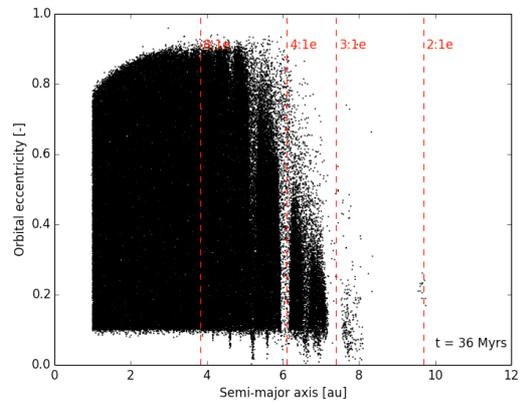
**Panel 4:** 36 Myr.

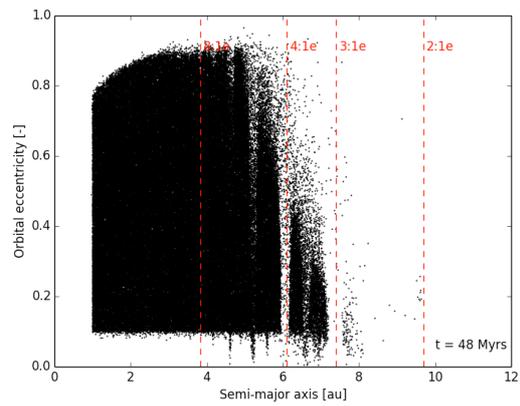
**Panel 5:** 48 Myr.

*Fig 2. The distribution of surviving test particles orbiting HR 8799 in our simulations, as a function of time, from t = 0 to t = 48 Myr, in semi-major axis-eccentricity space. Note the rapid initial dispersal of debris beyond ~8 au, followed by a more sedentary sculpting of the disk over the remaining duration of the integration.*

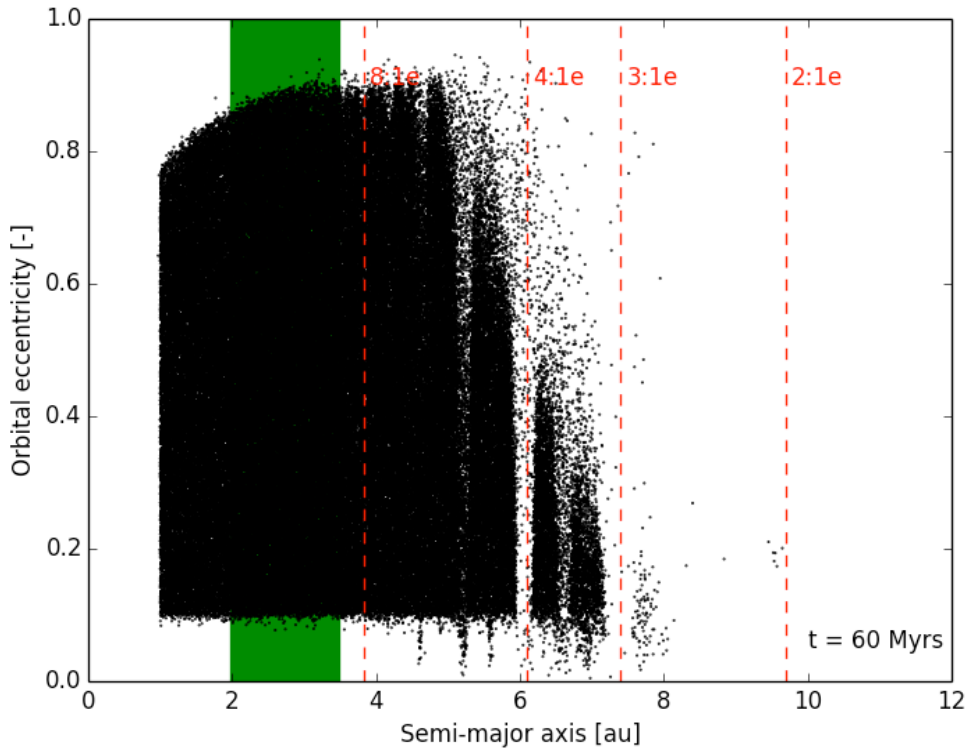

*Fig 3. The final distribution of test particles orbiting HR 8799 in semi-major axis – eccentricity space after 60 Myr. The sculpting influence of mean-motion resonances between test particles and planets can be clearly seen both in the gaps introduced to the disk (e.g. just outside 6 au), and in those locations where test particles have been driven to stable orbits with lower eccentricities (the downward "spikes" in eccentricity seen between 4-8 AU). The region shown in green (1.974 to 3.407 au) is our best estimate of the location of the Habitable Zone in the system.*

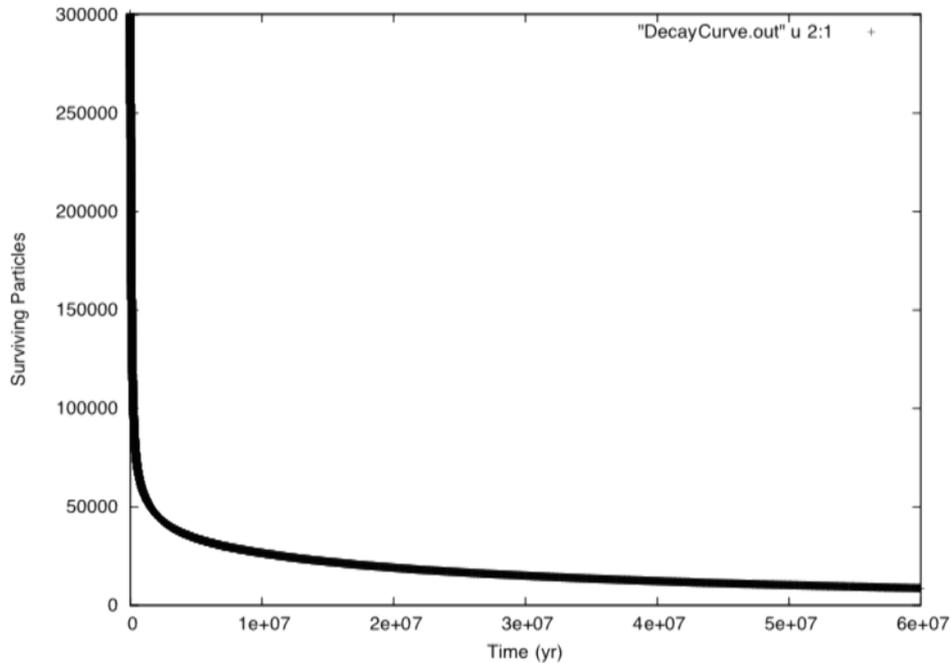

**Figure 4.** Plot showing the number of surviving test particles as a function of time. Unsurprisingly, given the extreme range of orbital eccentricities and semi-major axes tested in this work, almost 90% of the test particles are removed in the first 10 Myr of the disk's evolution.

## Discussion

Over the course of 60 Myr of dynamical evolution, the warm belt orbiting HR 8799 undergoes drastic sculpting, particularly during the first 6 Myr of integration (as seen in Figure 4). This fast clearance is the result of strong dynamical stirring caused by HR 8799 e, the innermost planet, which quickly clears almost all objects located exterior to the 3:1 mean-motion resonance with that planet. Apart from few objects that remain at ~ 9.7 au, probably trapped in the 2:1 MMR and also at ~ 8 au, by the end of the simulations, the outer edge of the disk has become clearly defined at ~ 7.5 au, revealing that test particles cannot survive beyond this distance unless trapped in mean motion resonance with one of the giant planets.

Furthermore, at the final time-step, the internal structure of the disk can be also observed. Several resonant features are evident, observed as cleared gaps within the belt, such as that centred on the location of the 4:1 mean motion resonance with HR 8799e. Other high order resonances with HR 8799e can be seen between 8:1e and 3:1e, in which the massless test particles were either cleaned or damped to low eccentricity orbits. These structural features are strongly reminiscent of the Kirkwood gaps, which are observed within the Asteroid belt in our own Solar system (e.g [28][29]). Interestingly, our results agree nicely with [19], who examined whether a hypothetical fifth planet could exist in the HR 8799 system, interior to the orbit of HR 8799e, in an orbit resonant with that planet (and thereby with the others in the system). They found that a hypothetical HR 8799f would be dynamically stable if it were to orbit at either $a \sim 9.7$ au or $a \sim 7.5$ au. In addition, in their figure 22, they map the stable regions for objects interior to HR 8799e. Their results, which were obtained through million-year MEGNO ([35]) integrations of the system, reassuringly reveal the same stable and unstable regions as those resulting from our longer n-body dynamical simulations.

In addition, a clear sculpting can be seen at high eccentricities at the inner edge of the disk. These highly eccentric objects move on orbits with small periastra and orbital periods such that the 7-day time-step used for these integrations is insufficient. This feature is therefore a computational effect, rather than the result of a real dynamical process, although it is exacerbated by our initial setup decision that no test particle should move on an orbit with periastron distance less than 0.1 au.

We also note that very little sculpting can be seen within the classical HZ. This suggests that that region might well be able to host dynamically stable Earth-like planets. Unfortunately, the HR8799 system is certainly too young to be a sensible target for the future search for life beyond the Solar system (e.g. [30]). However, the possibility that such exo-Earths could exist in a system that so closely resembles our own remains intriguing. In future work, we will build on earlier studies of the influence of giant planets on the impact flux in our own Solar system (e.g. [31]; [32]) to examine the potential impact regimes that such exoEarths in the system might experience. This will allow us to build tools that will eventually be used to help determine which Earth-like planets, discovered in the years to come, are the most promising targets for the search for life beyond the Solar system [31], allowing astronomers to infer both the impact regime that would be experienced by those planets, and the potential level of hydration that they might have received from exogenous sources.

## Conclusions

We have performed a detailed dynamical analysis of the inner planetesimal belt in the HR8799 planetary system. By simulating the orbital evolution of 500,000 massless test particles, initially distributed on dynamically hot orbits interior to the system's innermost planet, we have provided strong constraints on the structure and extent of the planetesimal belt.

Our results reveal that the outer edge of the inner belt can lie no further than ~7.5 au from its host star. The simulations also show that the belt must have a significant amount of internal structure, containing several gaps, cleared by the influence of mean-motion resonances with the giant planets in the system. These gaps, a direct analogue of the Solar system's Kirkwood gaps, are most clearly seen around the locations of the 4:1 and 3:1 mean-motion resonances with HR8799 e. Exterior to the location of that 3:1 mean-motion resonance, a small population of test particles can be seen trapped between ~ 7.5 to 8 au, at low eccentricities.

Complementary studies are required in order to investigate the evolution of a dynamically cold debris disk (the low eccentricity, low inclination counterpart to the disk studied in this work). Those simulations have recently been completed, and the results will be presented in a future paper. We will examine the impact rates that might be expected on any planets located in the HZ, as well as studying the possible routes by which water, and other volatiles, might be delivered to those planets. Further work will yield a better understanding of the dynamical structure of the inner debris belt, and offer clues as to the possibility and viability of a fifth, terrestrial planet in the HZ.

## Acknowledgement
This work is supported by CAPES Foundation, Ministry of Education of Brazil, Brasilia – DF, Zip Code 70.040-020 and School of Physics - UNSW Australia, and made use of UNSW Australia's *Katana* supercomputing cluster.